\documentclass[final,1p,number,sort&compress]{elsarticle}

\usepackage{amssymb}

\usepackage{lineno}

\journal{Nuclear Instruments and Methods A}

\begin{document}


\begin{frontmatter}



\title{Studies of relative gain and timing response of fine-mesh 
  photomultiplier tubes in high magnetic fields}


\author[address_1]{V.~Sulkosky\corref{cor1}}
\cortext[cor1]{Corresponding author. Tel.: +1-434-924-7589; fax: +1-434-924-7909.}
\ead{vasulk@jlab.org}
\author[address_2]{L.~Allison}
\author[address_3]{C.~Barber}
\author[address_3]{T.~Cao}
\author[address_3]{Y.~Ilieva}
\author[address_1]{K.~Jin}
\author[address_2]{G.~Kalicy}
\author[address_2]{K.~Park}
\author[address_1]{N.~Ton}
\author[address_1]{X.~Zheng}

\address[address_1]{University of Virginia, Charlottesville, VA, 22904, USA}
\address[address_2]{Old Dominion University, Norfolk, VA, 23508, USA}
\address[address_3]{University of South Carolina, Columbia, SC, 29208, USA}

\begin{abstract}
We investigated the use of Hamamatsu fine-mesh photomultiplier tube assemblies
H6152-70 and H6614-70 with regards to their gain and timing resolution in 
magnetic fields up to 1.9~T.  Our results show that the H6614-70 assembly can 
operate reliably in magnetic fields exceeding 1.5~T, while preserving a 
reasonable timing resolution even with a gain reduction of a factor of 
$\approx$ 100.  The reduction of the relative gain of the H6152-70 is similar 
to the H6614-70's near 1.5~T, but its timing resolution worsens considerably 
at this high field. 

\end{abstract}


\end{frontmatter}


\section{Introduction}
\label{sec:intro}
In nuclear and particle physics experiments, photomultiplier tubes (PMTs) are commonly used to 
collect photon signals from scintillating detectors.  Many such detectors are used for timing 
measurements, such as time of flight (TOF), which require not only a large signal amplitude, but 
also a good timing resolution.  The timing resolution achieved by the scintillator is directly related 
to the number of photoelectrons emitted by the PMT's cathode that reduces statistically the timing fluctuation of 
the PMT's signal due to the scintillator's decay time, the propagation of photons through the scintillator, 
and the time for the photoelectron(s) to transit from the PMT's cathode to the anode. The performance
of scintillators with a PMT readout is highly sensitive to the presence of external magnetic fields 
because of the field effect on the photoelectrons.  For typical PMTs, the trajectory of the 
photoelectron between the PMT's photocathode and the first dynode is affected by fields as 
low as a few Gauss, causing a significant loss in statistics and a larger timing fluctuation 
in the PMT's output~\cite{Bonesini:2012wu}.  The use of mu-metal 
shielding can extend the operating range of PMTs up to $\approx$ 100~Gauss.  However, 
detectors used in modern experiments are often part of large spectrometer systems that include 
one or more magnets with fields at the Tesla level. To meet the need for operation in Tesla-level 
fields, special PMTs with fine-mesh type dynodes have been developed and put into 
use~\cite{Enomoto:1993xz,Iijima:1996uc}.

At the Thomas Jefferson National Accelerator Facility (JLab), the Solenoid Large Intensity Detector 
(SoLID)~\cite{Chen:2014psa} is being designed to be a large acceptance and a high luminosity device 
in experimental Hall A.  This device is a multi-purpose spectrometer to study physics topics such as 
semi-inclusive deep inelastic scattering (SIDIS) from polarized targets, threshold $J/\psi$ production, 
and parity-violating deep inelastic scattering (PVDIS).  The SoLID apparatus consists of a solenoid with 
a magnetic field of approximately 1.5~T and an open-geometry detector package.  The PVDIS configuration 
will consist of one large-angle detector package, while the SIDIS-$J/\psi$ configuration will have two 
detector packages: one at forward angles and one at large angles.  The detector packages for SIDIS will 
include a set of GEM detectors for tracking, a scintillator pad 
detector (SPD) for TOF measurements and for trigger-rate reduction by photon rejection, a Multi-Gap Resistive Plate 
Chamber (MRPC) to provide a TOF measurement at forward angles, a light gas Cherenkov counter to identify electrons, 
a heavy gas Cherenkov counter for hadron identification, and an electromagnetic calorimeter (EC) for electron 
identification.  For PVDIS, the heavy gas Cherenkov and SPD are not used, otherwise the detector package is similar 
to that of SIDIS.  TOF capability is a critical requirement for the SIDIS experiments.  At forward angles, 
TOF is provided by the MRPC, while at large angles (LA), the LASPD is the only TOF detector.  The LASPD must provide 
time-of-flight resolution of 150 ps or better, which requires fast, high-photon yield scintillators and fast PMTs.  
To achieve the required timing resolution, the PMTs need to be attached directly to the 
LASPD bars, which means that the PMTs must operate inside the solenoid field of 1.5~T.  

One possible PMT choice for the LASPD is a fine-mesh (FM) PMT with a large effective
area, high gain, and small timing jitter.  The high-field resistance of these PMTs is achieved by using a 
fine-mesh dynode structure with layer separation of $\approx$~1~mm.  The first dynode is 
a few millimeters from the photocathode, and this small distance allows for efficient collection and multiplication 
of the primary photoelectrons even in the presence of a high magnetic field.  Fine-mesh PMTs date back to the early 
1980s and the early designs were based on the studies conducted at that time~\cite{Takasaki:1985}.  Since these early 
prototypes, the FM-PMT design has matured with various studies conducted over the past few  
decades~\cite{Takasaki:1985,Enomoto:1993xz,Iijima:1996uc,Bonesini:2006zz,Bonesini:2007zz,Boeglin:2008,Bonesini:2012wu}.
Some of these investigations involved the application to various detector types such as threshold Cherenkov 
counters~\cite{Iijima:1996uc}, which typically have small light yields.  These studies aided in the 
optimization of the number of dynode stages and the mesh spacing, improved magnetic field immunity 
and absolute gains.  However, most of the reported results were limited to magnetic fields less than 
1.2~T with only a few measurements up to 1.5~T that only included results on gain and pulse height resolution.  
The angular dependency of the earlier measurements was also limited for orientation angles with respect 
to the magnetic field within $\left(0^{\circ}, 90^{\circ}\right)$.  For the SoLID SIDIS program, the available 
data were insufficient to determine whether FM PMTs would operate reliably and provide timing resolution 
at the 150~ps level.

We have conducted studies of fine-mesh PMTs' performance in fields up to 1.9~T.  
The experimental setup is described in Section~\ref{sec:expsetup}.  In Section~\ref{sec:calibration}, various 
experimental and data analysis procedures are discussed.  In Section~\ref{sec:exp_results}, results 
on the FM PMTs' relative amplitude and timing resolution in magnetic fields up to 1.9~T for a range of 
orientation angles are presented.  Section~\ref{sec:conclusions} provides a summary of our study.

\section{Experimental setup}
\label{sec:expsetup}
All measurements involving magnetic fields presented here were performed in July~2015 
using the Jefferson~Lab High-B Sensor-Testing Facility in collaboration with Jefferson Lab, 
Old Dominion University, and the University of South Carolina.  The test facility was designed 
for gain evaluation of small photon sensors in magnetic fields up to 5~T.  Additional measurements to 
characterize the FM PMTs without a magnetic field were conducted at the University of Virginia (UVA) 
in June and September~2015.

\subsection{Fine-mesh photomultiplier tubes}
\label{sec:fm-pmts}
\begin{figure}[!ht]
\begin{center}
\includegraphics[width=11cm]{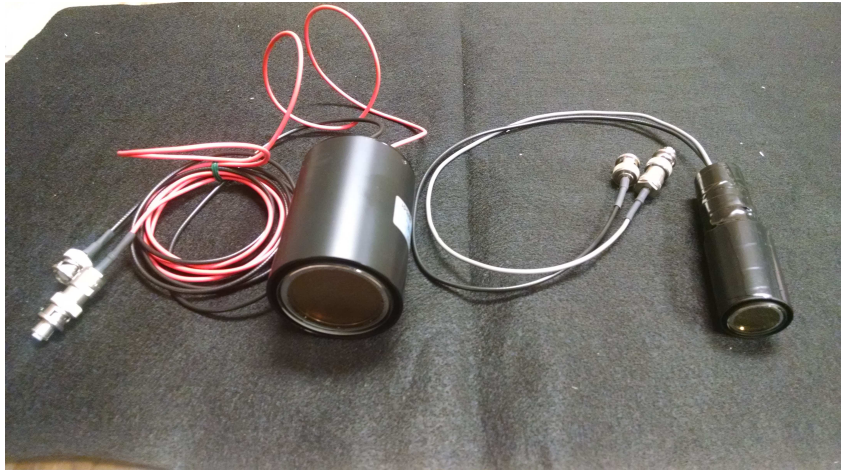}
\caption{The Hamamatsu H6614-70 (left) and H6152-70 (right) FM PMT assemblies.
  \label{fig:fmpmts}}
\end{center}
\end{figure}

The FM PMTs tested in our study are the R5505-70 ($\phi$ 25 mm) and the R5924-70 ($\phi$ 51 mm)
from Hamamatsu Photonics~\cite{Hamamatsu:2015}.  The main properties of these phototubes are presented 
in Table~\ref{table_pmts}.    
\begin{table}[!ht]
\caption{Relevant properties of tested Hamamatsu fine-mesh PMTs.  TTS stands for the transit time 
spread.\label{table_pmts}}
\begin{center}
\begin{tabular}{llll}
\hline
PMT               &R5505-70        &R5924-70  \\ 
\hline
Assembly          &H6152-70        &H6614-70  \\
Diameter (mm)     &25              &51        \\ 
Number of stages  &15              &19        \\
Risetime (ns)     &1.5             &2.5       \\
Transit time (ns) &5.6             &9.5       \\
TTS (ns)          &0.35            &0.44      \\
Gain at +2 kV     &$5 \times 10^5$ &$1 \times 10^7$ \\
\hline
\end{tabular}
\end{center}
\end{table}
Both PMTs use a bialkali photocathode with effective diameters of 17.5~mm and 39~mm, respectively.  The 
R5924-70 PMT was delivered as an assembly module H6614-70 without a $\mu$-metal shield.  The 
H6152-70 module was assembled with the R5505-70 PMT, the voltage divider E6133-04 MOD and a 
1~mm-thick $\mu$-metal shield, all from Hamamatsu.  The effectiveness of magnetic shielding using 
$\mu$-metal is limited due to saturation effects at $\approx$ 50--100~Gauss~\cite{Bonesini:2012wu}.  
Therefore, since the FM PMTs are tested over a range of fields extending far above 0.1~T, we do not 
expect the mu-shielding to have much effect on our measurements at the high end of the field range.  
The results we report here are from tests of one module each of the H6152-70 and H6614-70 assemblies.  
We operated the modules at a high voltage of +2 kV.  A picture of the two assemblies are shown in
Fig.~\ref{fig:fmpmts}.

\subsection{High field magnet}
\label{sec:magnet}
The superconducting solenoid~\cite{Keith:2012ad} of Jefferson Lab's High-B Sensor-Testing Facility 
has a warm bore with length of 76.2~cm and diameter of 12.7~cm.  The magnet can reach 5.1~T at 82.8~A,
but for this test we only went up to 1.9~T.  The central field inhomogeneity is less than
$5 \times 10^{-5}$ over a cylindrical volume that is 5~cm long with a diameter of 
1.5~cm.  During the measurement period, the magnet was manually refilled with liquid helium from a nearby dewar 
about every other day.

\subsection{Dark box and PMT holder}
\label{sec:darkbox}
\begin{figure}[!ht]
\begin{center}
\includegraphics[height=6cm]{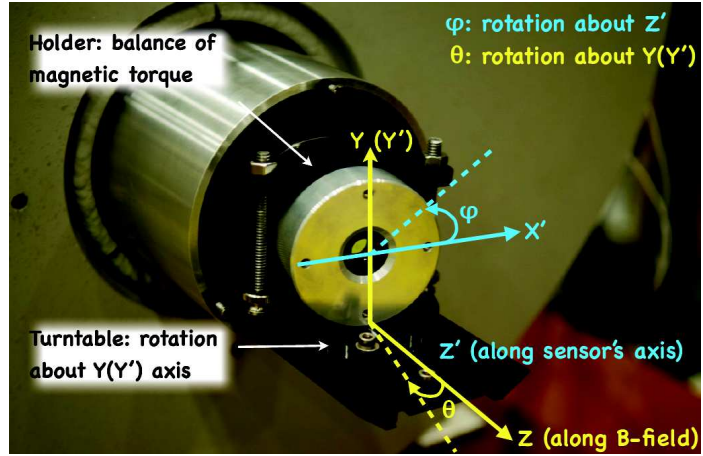}
\caption{PMT inside the dark box holder.  Axes are labeled to indicate the orientation
  of $\theta$ and $\phi$ with respect to the magnetic field direction.  Z is along the 
  magnetic field direction, X (X') is horizontal, and Y (Y') is vertically up.   When 
  the PMT is rotated away from $\theta$ = 0$^{\circ}$, Z' is along the PMT's axis.
  \label{fig:orientation}}
\end{center}
\end{figure}

For the tests presented here, a cylindrical dark box of diameter 11.4~cm and a length of 45.7~cm
was used to hold the PMTs.  All components of the dark box were non-magnetic.  
The PMTs were placed one at a time inside the dark box and were held firmly in place with a 
holder (see Fig.~\ref{fig:orientation}) to balance the magnetic torque in case any 
component of the PMT assembly was magnetic.  Figure~\ref{fig:orientation} shows a picture of a test 
PMT inside the holder along with definitions of the polar and the azimuthal angles $\theta$ and $\phi$, 
respectively, used to describe the orientation of the PMT with respect to the field.  Angle $\theta$ is 
the angle between the PMT and the field axes, and $\phi$ is the azimuthal angle of the PMT with respect 
to an axis perpendicular to the PMT's axis.
The angle $\theta$ can be adjusted in 5$^{\circ}$ increments up to $\pm$~60$^{\circ}$ by using a turntable; 
a positive angle indicates a clockwise rotation about Y (Y'), and a negative angle a counterclockwise rotation.    
When the angle was changed, the new position was locked into place by tightening a screw into the turntable 
assembly.  For angle $\phi$, the orientation was determined by markings on the PMTs outer casing.  Due to the 
imprecision of locating the markings, only rotations in increments of approximately 90$^{\circ}$ were attempted.  
The dark box was positioned and centered inside the solenoid's bore by utilizing marks made on the bore's surface, 
which were previously measured.

The dark box's light tightness was achieved with two endcaps located at each end of the
cylinder.  One endcap contained a connector located at its center to which a 5-m long
optical fiber was screwed into place.  The optical fiber was used to transport light produced by a 
light-emitting diode (LED) into the dark box.  On the inside of the cap, a diffuser was attached to 
illuminate the entire surface of the PMT inside the dark box.  The other endcap included SHV and BNC 
feed-through connectors to accommodate the high-voltage and the signal cables.  To avoid any potential 
light leaks, the magnet's bore openings were covered with black material to keep ambient light from 
entering the bore.

\subsection{Light-Emitting Diode (LED)}
\label{sec:LED}
During the measurements at Jefferson Lab, an LED\footnote{THORLABS LED405E} with a central 
wavelength of 405~nm (FWHM = 15 nm) was used as a photon source.  An HP 8116A pulse generator 
was used to drive the LED with a 10-ns wide square pulse, an adjustable amplitude up to 8~V, and a 
frequency of 30~kHz.  The measurements of the H6152-70 and H6614-70 asssemblies were conducted with 
a pulse-generator amplitude of 7.5 V and 7.0 V, respectively.  These values were optimized by scanning
the PMT timing resolution over the range of amplitudes between 4.8 and 8 V and were selected based on
the amplitude providing the best resolution.  The light from the LED was split equally into a bifurcated  
50-$\mu$m diameter fiber with legs of length 2~m; one of these legs was sent to a SiPM\footnote{Size: 3$\times$3~mm$^2$, 
3600 50~$\mu$m pixels} that was used as a light monitor, and the second leg was connected to the 5-m long optical 
fiber to illuminate the PMT inside the dark box.

\subsection{Data acquisition}
\label{sec:DAQ}
\begin{figure}[!ht]
\begin{center}
\includegraphics[height=6cm]{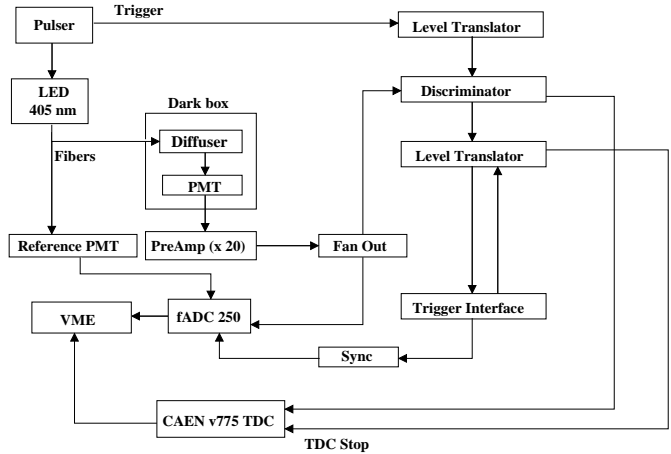}
\caption{Circuit diagram of electronics used for gain and timing resolution measurements.  In the text,
  the reference PMT is referred to as the SiPM.
  \label{fig:circuit}}
\end{center}
\end{figure}

A diagram of the PMTs' readout and the LED triggering is shown in Fig.~\ref{fig:circuit}.
The trigger output signal from the pulse generator was used to trigger the data acquisition (DAQ) system.  
The signal from the test PMT was fanned-out and sent to a CAEN v775 Time-to-Digital Converter (TDC) 
and a F250 flash Analog-to-Digital Converter (fADC) custom-made by Jefferson Lab~\cite{Dong:2007}.  
A leading edge Phillips Scientific (PS) 706 discriminator with a minimum threshold of $-$10~mV was 
used to provide the START signal to the TDC.  A 20$\times$ fast preamplifier (ORTEC VT120 version C) 
boosted the signal so that the PMT pulses would pass the discriminator threshold at higher magnetic 
fields.  The trigger signal was also sent to the TDC to provide a reference time 
and to remove the 4~ns jitter caused by the trigger interface (TI) module, which generated the signal 
used for the TDC STOP.  The fADC (1.0~V full scale) sampled the PMT output 
voltage every 4~ns, and 250 samples were stored for each event.  The DAQ software CODA~\cite{Abbott-CODA:2009}
 (CEBAF Online Data Acquisition) version 2.5 was used to record the digitized PMT pulses from the 
three modules.

\section{Calibration and normalization}
\label{sec:calibration}

\subsection{TDC resolution}
\label{sec:tdcres}
Prior to beginning the measurements with a magnetic field, the TDC module was calibrated to 
verify its intrinsic resolution.  The TDC's full-scale range ($\approx$ 140~ns) was set 
to provide an intrinsic timing resolution of 35~ps per bin.  
The TDC was calibrated by solely using the trigger output signal of the pulse generator.  The 
signal was split, with one branch providing the TDC STOP signal and the other branch used as a 
TDC input signal.  The branch for the STOP was first sent through the TI module to trigger the DAQ,
which produced a delay of $\approx$ 180 ns between both branches when they arrived at the TDC.
In order for the TDC input signal to arrive within the 140~ns window, the input signal was delayed
by 51~ns with a PS 792 NIM module.  Then additional delays of the TDC input signal were provided by 
this delay module. For each delay, the position of the corresponding timing peak was measured by 
the TDC.  The TDC calibration was then derived from the correlation between the time delay and the 
TDC peak position.  The average TDC intrinsic resolution was determined to be 34.1 $\pm$ 0.4~ps/bin.

\subsection{Signal processing of fADC data}
\label{sec:fADC_data}
\begin{figure}[!ht]
  \begin{center}
    \includegraphics[width=0.99\textwidth]{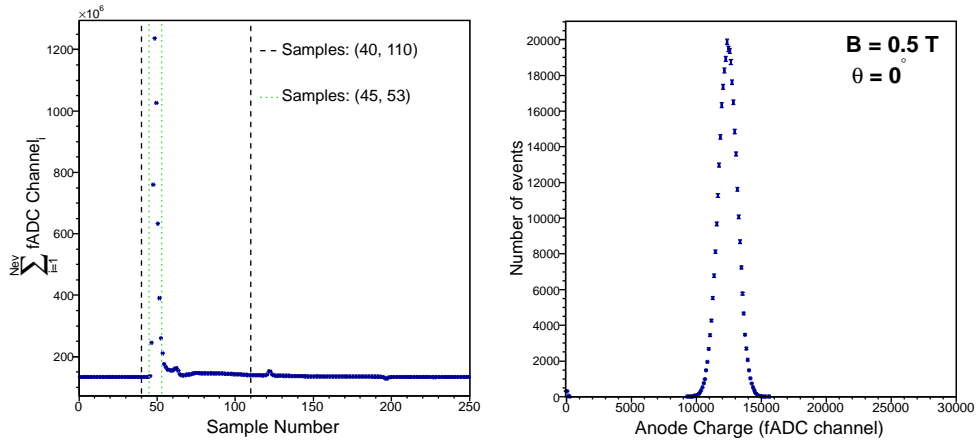}
    \caption{The left-side panel shows the fADC cumulative waveform measured for 
      $\approx 7 \times 10^{5}$ events at 0.5~T and 0$^{\circ}$.  The
      cumulative waveform visualizes the shape of the average waveform of all
      the events and was used as a monitoring tool during data taking.  For
      each setting (B, $\theta$), the baseline (pedestal) is determined as the
      average of the first 35 samples.  The right-side panel shows the event
      distribution of the pedestal-subtracted, integrated charge collected on
      the anode in units of fADC channel.  For each event, this quantity was
      obtained as the sum of the recorded fADC digits over a subrange of
      samples (40, 110), as depicted by the dashed lines in the left-panel,
      which is equivalent to integrating the signal waveform over that subrange.
      In the later sections of this paper, we refer to the mean value of the 
      integrated-charge distribution as the ``amplitude''.
      \label{fig:fadc_sample}}
  \end{center}
\end{figure}

Figure~\ref{fig:fadc_sample} (left-panel) illustrates an example of the fADC read out samples 
from the H6614-70 assembly; the histogram represents the cumulative waveform of $\approx 7 \times 10^{5}$ 
events at 0.5~T and $\theta$ = 0$^{\circ}$.
The fADC baseline (pedestal) was measured for each $\theta$ at zero 
magnetic field with the pulse generator disabled (LED off) and was determined by averaging the 250 
samples of the fADC distribution.   These measurements yielded a sample of 95 pedestal values.  The 
standard deviation of that sample was less than 0.5\% of the sample mean, indicating that the fADC 
response was stable to this level over the measurement period.   To remove the effect of any instability 
of the fADC pedestal over the time period of the measurements, the baseline was also determined directly 
from the data when the LED was on by averaging the first 35 fADC samples.  The right-side panel of 
Fig.~\ref{fig:fadc_sample} represents the pedestal-subtracted integrated charge, which was obtained 
by summing the fADC digits for each recorded event over a subrange of samples (40, 110).  At each setting 
(B, $\theta$), the mean value of the integrated-charge distribution is proportional to the gain of the 
FM PMT at that setting, and we used it to quantify the gain performance of the sensor as a function of B 
and $\theta$.  In the later sections of this paper, we refer to this mean value as the ``amplitude''.

Events associated with the pedestal were removed by requiring a valid hit in 
the TDC.  At magnetic fields above 1~T, where the signal amplitudes are very low and thus, very close 
to the pedestal, this requirement is critical to cleanly separate the signal from the pedestal.  A 
narrower integration range over 9 samples (dashed-dotted line) was used to estimate the systematic 
uncertainty due to the chosen sample range.  For the two subranges, the difference was typically better 
than 5.5\%; however, for magnetic fields less than 1~T, the difference was $\sim$ 1\%.  Above 1~T, the 
signal-to-noise ratio worsens significantly, which causes the larger differences at these 
fields.  The results presented in Section~\ref{sec:exp_results} are from the analysis using the subrange 
of (40, 110) samples.

\subsection{Timing resolution dependence on the orientation angle at 0~T}
\label{sec:tres_orientang}
The timing resolution of time-of-flight detectors can be parameterized~\cite{Bonesini:2012wu} 
as
\begin{eqnarray}
  \sigma_{T} &=&  \sqrt{\frac{\sigma_{TTS}^2 + \tau_{sct}^2 
      + \sigma_{pl}^2}{N_{PE}} + \sigma_{elec}^2}\>,
  \label{sigmat.eq1}
\end{eqnarray}
where $\sigma_{TTS}$ is the transit time spread of the PMT (see Table~\ref{table_pmts}), $\tau_{sct}$ is 
the light decay constant of the scintillator, $\sigma_{pl}$ is caused by path-length variations, 
and $\sigma_{elec}$ is the jitter of the electronics readout.  Finally, $N_{PE}$ is the average number 
of photoelectrons detected.  In the test done at Jefferson Lab, $\tau_{sct}$ should be replaced 
with $\tau_{LED}$, which reflects the exponential decay of the LED intensity caused by the discharge of 
a capacitor, and the $\sigma_{pl}$ term is negligible because a scintillator was not involved and 
the distance between the light source and PMT surface is approximately the same across the PMT's 
photocathode.    The $\tau_{LED}$ term can be minimized by using an inductance in parallel to the 
LED~\cite{Kapustinsky:1985,Lubsandorzhiev:2004zh}.  However, this was not implemented in the test 
facility at Jefferson Lab.   The term $\tau_{LED}$ was empirically determined to be $\sim$ 2~ns from 
the data, which dominates $\sigma_{T}$ in our measurements.  The absolute timing resolution for the 
H6614-70 assembly was limited to $\approx$ 280~ps with a pulse generator amplitude of 7.0~V at 0~T and 
$\theta$ = 0$^{\circ}$ due to the large $\tau_{LED}$.  The preamplifier added $\sim$ 20--30~ps to this 
resolution.  

\begin{figure}[!ht]
  \begin{center}
    \includegraphics[width=0.85\textwidth]{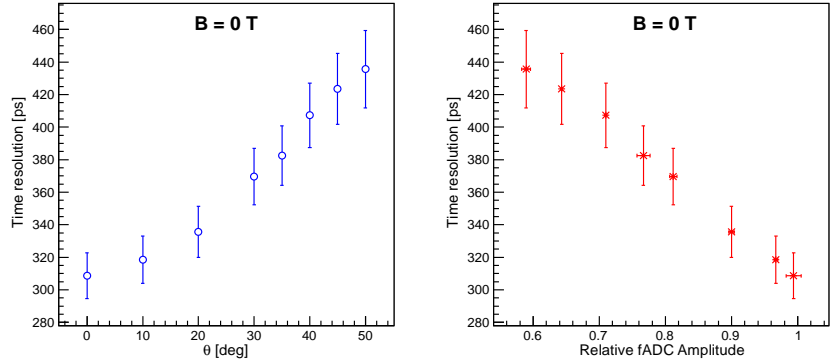}
    \caption{The measured timing resolution of the H6614-70 assembly at 0~T versus the relative orientation angle 
      of the PMT (left panel) and the timing resolution versus relative amplitude (right panel).  The fADC 
      amplitudes were normalized to the results at 0$^{\circ}$.
      \label{fig:RESVSANG}}
  \end{center}
\end{figure}

One of the limitations of the test facility is that the optical fiber was fixed in place at the front of 
the dark box as the PMT's angle was rotated with respect to the magnetic field; this feature caused the 
amount of light reaching the PMT's photocathode and hence $N_{PE}$ to decrease with larger 
relative orientation angles in the absence of the magnetic field.  Unfortunately, this feature of 
the system was not easy to change due to the small bore of the solenoid.   For each $\theta$, data 
were taken at zero magnetic field in order to correct for the loss of $N_{PE}$ as the angle increased 
from $0^{\circ}$.  Figure~\ref{fig:RESVSANG} shows the measured timing resolution versus $\theta$ in 
the left panel.
The resolution worsens by about 40\% from $0^{\circ}$ to $50^{\circ}$.  The error bars on the data points reflect 
an estimate of the average $N_{PE}$ detected.  In the right panel, the timing resolution versus relative fADC 
amplitude is plotted, where the amplitudes have been normalized to the data at $0^{\circ}$ and 0~T.  The horizontal 
error bars represent the standard deviation of measurements at the same angle, and the amplitude is proportional 
to the average number of photoelectrons.

\subsection{Estimation of number of photoelectrons}
\label{sec:est_npes}
The single photoelectron (SPE) response was not seen in the measurements conducted at Jefferson Lab.
After these tests concluded, the SPE was measured at UVA.  Based on these data, 
the estimated number of photoelectrons measured by the FM PMTs at zero magnetic field is 66$^{+14}_{-10}$ 
at $0^{\circ}$ and 39$^{+9}_{-6}$ at $50^{\circ}$ for the H6614-70, and 27$^{+7}_{-4}$ at $0^{\circ}$ for the 
H6152-70.  The large uncertainties are due to the difficulty in isolating the single 
photoelectron peak for FM PMTs~\cite{Enomoto:1993xz} and to the systematic uncertainties involved in combining 
the data taken at Jefferson~Lab and UVA.

\section{Experimental Results}
\label{sec:exp_results}
\begin{figure}[hbtp]
\begin{center}
\includegraphics[height=7cm]{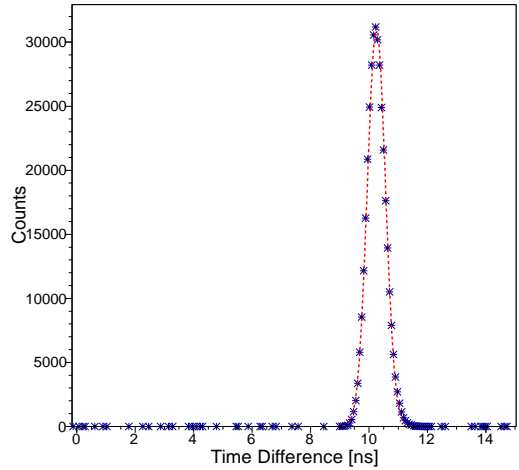}
\caption{The time difference between the PMT time and the trigger time at 0.5~T and $0^{\circ}$ as 
  measured by the TDC in ns.  The dashed line represents a Gaussian fit to the data, giving 
  $\sigma_{T}$ = 311~ps.
  \label{fig:tdiff}}
\end{center}
\end{figure}

Measurements were completed to determine the relative gain reduction and timing resolution as a function
of magnetic field and orientation angle for the two types of FM PMTs. 
The timing resolution was determined by fitting the time difference between the PMT time and the trigger
time as measured by the TDC.  By taking the time difference, the 4~ns jitter induced by the trigger interface
module was removed.    Corrections for the time-walk effect were attempted but found to typically improve 
the timing resolution by less than 5~ps.  Compared to the measured timing resolutions of 300$-$400~ps, these 
corrections were small and hence not included in the final analysis.  An example time-difference distribution
at 0.5~T and $0^{\circ}$ is shown in Fig.~\ref{fig:tdiff} along with a fit to the data (dashed line).

\subsection{Two inch PMT assembly (H6614-70)}
\label{sec:2in_results}
\begin{figure}[hbtp]
  \begin{center}
    \includegraphics[width=0.49\textwidth]{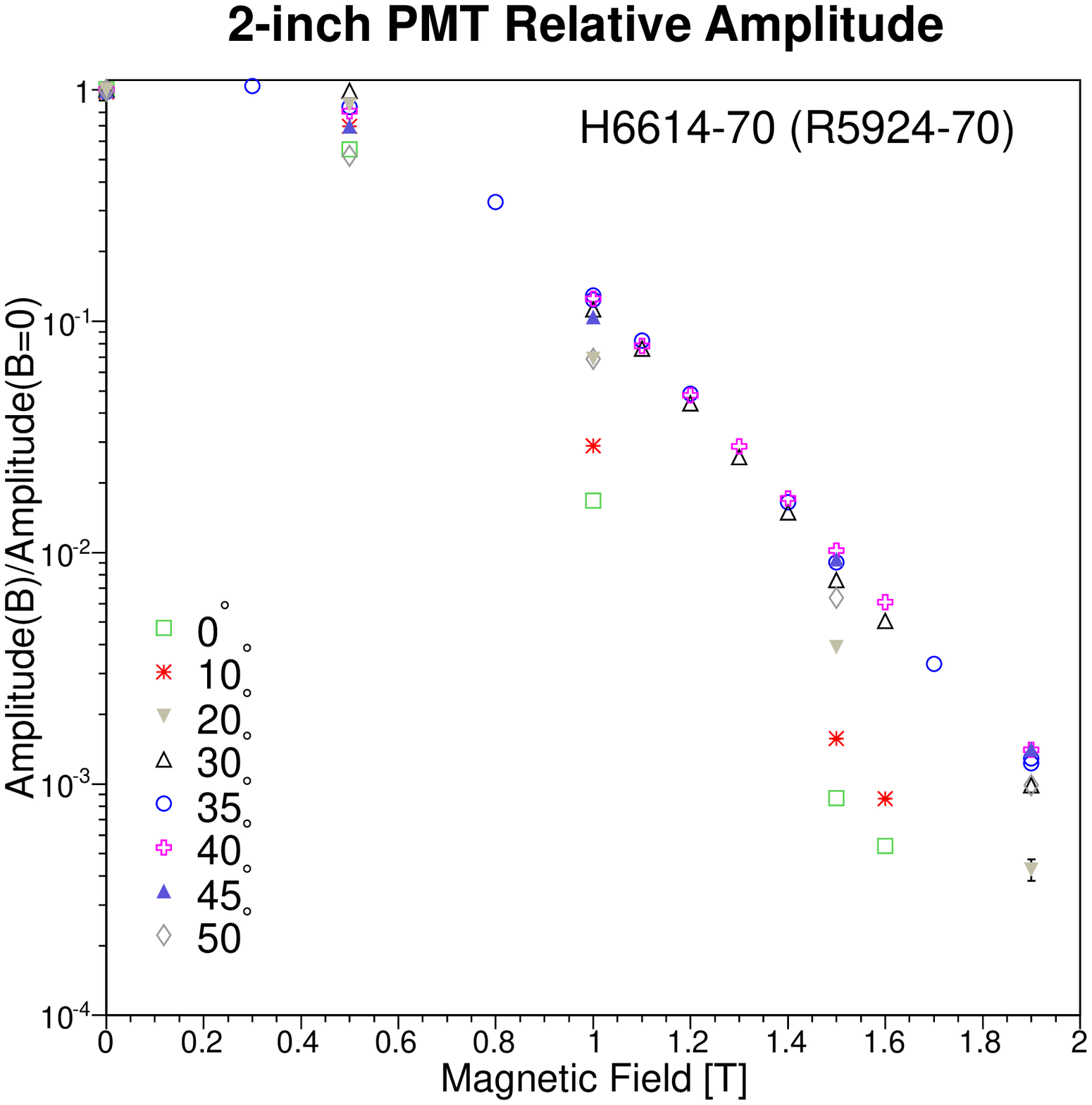}
    \includegraphics[width=0.49\textwidth]{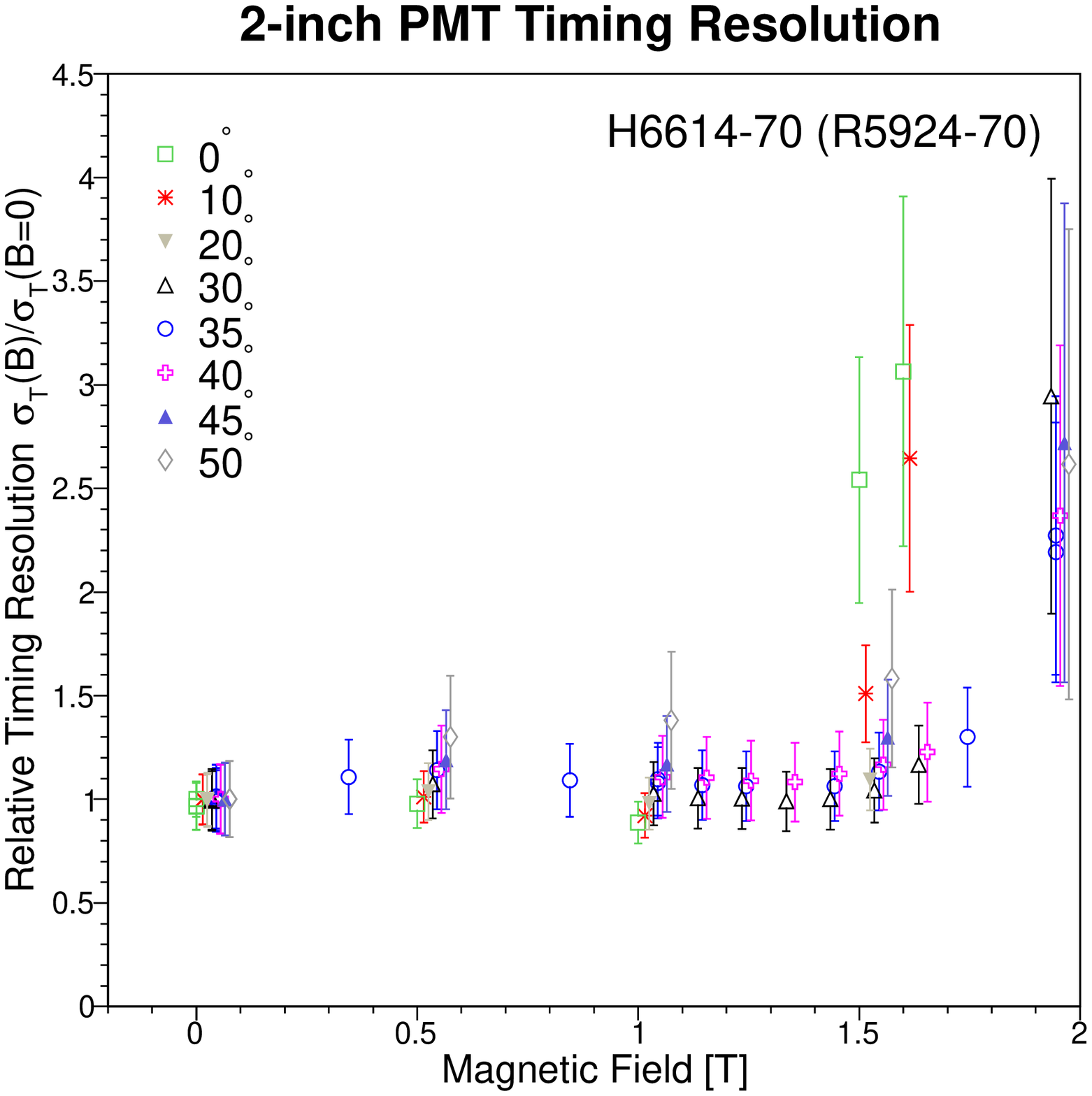}
    \caption{(Left) Relative amplitude versus magnetic field for the 2 in.\ Hamamatsu H6614-70 assembly 
      at various orientation angles, $\theta$.  The amplitudes are normalized to the values at 0~T and at 
      the same $\theta$ to determine their relative change as the field is increased.  (Right) The 
      relative timing resolution versus magnetic field.  A small horizontal shift was 
      added to the data at the same magnetic field so that the points are easier to see.
      \label{fig:H6614-70_all}}
  \end{center}
\end{figure}

For the H6614-70 (2~in.)~assembly, measurements were conducted for orientation angles from 0$^{\circ}$
up to 50$^{\circ}$ with magnetic fields from 0 to 1.9~T.  At 0$^{\circ}$ and 10$^{\circ}$, tests
were conducted only up to 1.6~T due to the lack of signal at higher fields. 
The reduction in relative amplitude (gain) is illustrated in the left-side panel of 
Fig.~\ref{fig:H6614-70_all}, where the amplitudes are normalized to the amplitude at 0~T at the 
same angle.  Statistical uncertainties are included on the points; however in most cases, they cannot 
be seen due to the large number of events.  At each angle, data were recorded at 0~T before 
ramping the field up and again after the measurements were completed and the field was ramped down.  The 
amplitudes before and after ramping were consistent with each other to better than 2\%.  As expected, 
a reduction in amplitude occurs with increasing {\bf B}-field magnitude, and the amplitude shows a strong 
dependence with $\theta$.  For $\theta$ = 35$-$45$^{\circ}$, the relative amplitudes are approximately 
the same with a reduction of $\sim$ 100 at 1.5~T.  Below 20$^{\circ}$ the reduction factor approaches 
1000 at {\bf B} = 1.5~T.

The right-side panel of Fig.~\ref{fig:H6614-70_all} shows the relative timing resolution ($RT$) 
for the same range of orientation angle and magnetic field.  The error bars represent the 
uncertainty as determined from an estimation of the number of photoelectrons with 
$\delta \sigma_{T}/\sigma_{T} \propto 1/\sqrt{N_{PE}}$.  
The values for $\sigma_{T}$ were normalized to the results at $\theta$ = 0$^{\circ}$ and zero {\bf B} field 
so that $RT = \sigma_{T}\left(\theta, \mathrm{B}\right)/\sigma_{T}\left(0^{\circ}, \mathrm{B} = 0\right)$
and then multiplied by $\sigma_{T}\left(0^{\circ}, \mathrm{B} = 0\right)/\sigma_{T}\left(\theta, \mathrm{B} = 0\right)$
to correct for the loss of $N_{PE}$ as $\theta$ is rotated away from 0$^{\circ}$.   
For angles between 20$^{\circ}$ and 45$^{\circ}$, as the magnetic field increases from zero~T, a small worsening 
of the resolution is seen, then the resolution is approximately flat between 0.5~T and 1.5~T.  The resolution 
then becomes significantly worse at 1.9~T, suggesting a loss of $N_{PE}$ between the photocathode and the 
first dynode.

\begin{figure}[hbtp]
  \begin{center}
    \includegraphics[width=0.49\textwidth]{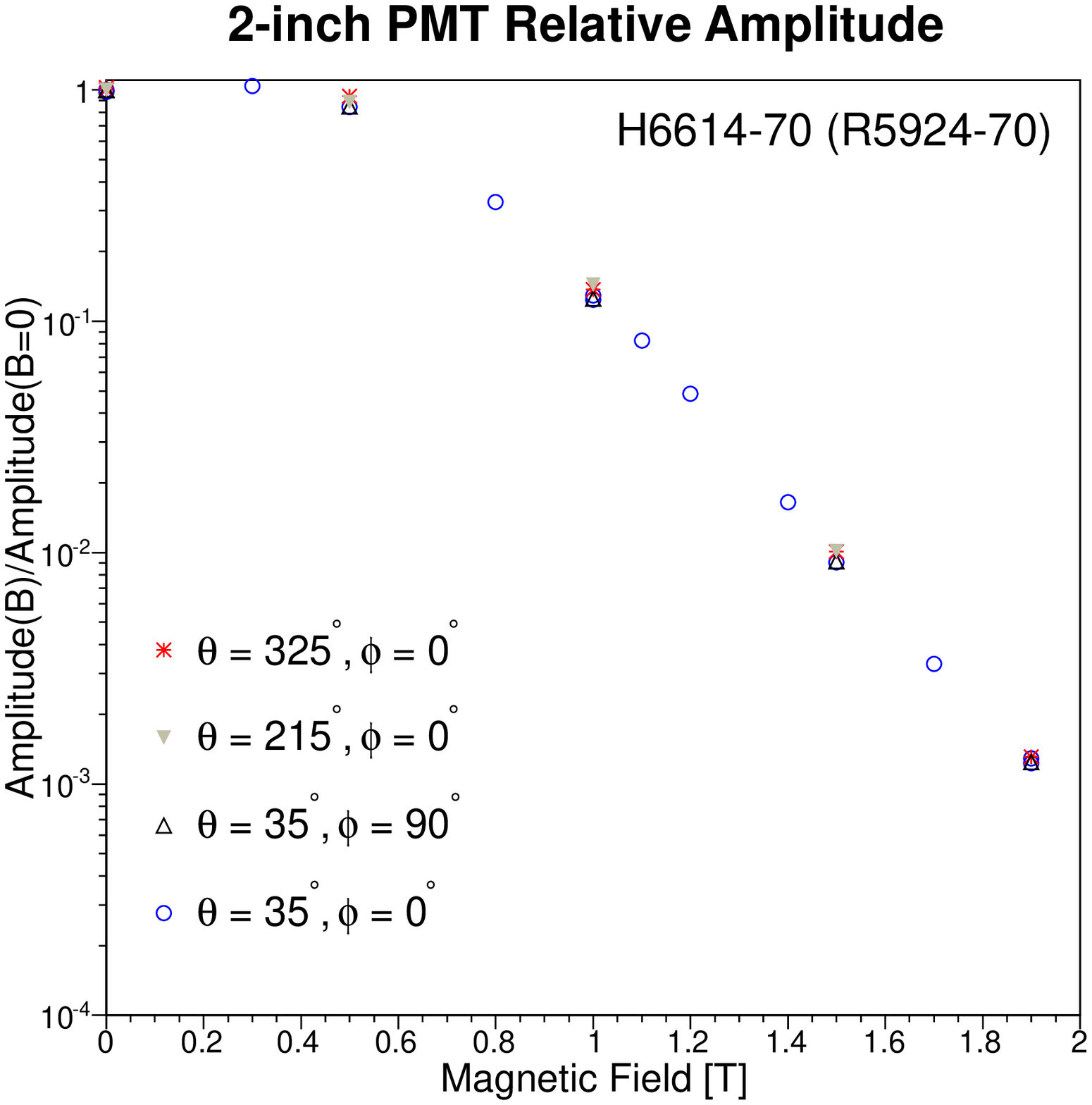}
    \includegraphics[width=0.49\textwidth]{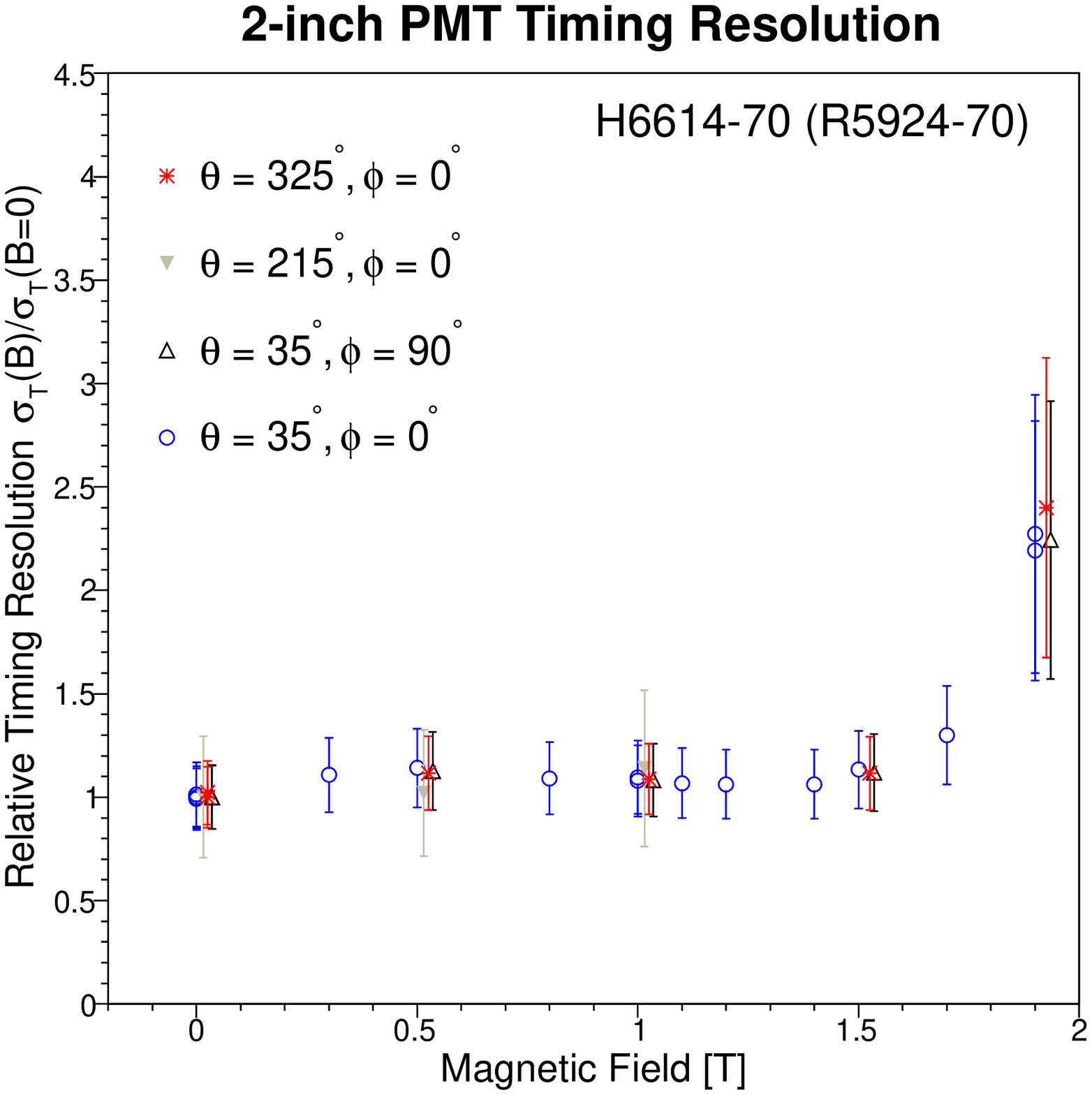}
    \caption{(Left) Relative amplitude versus magnetic field for the 2~in.\ Hamamatsu H6614-70 assembly 
      at $\theta$ = 35$^{\circ}$, 215$^{\circ}$, and 325$^{\circ}$ with azimuthal angles at
      $\phi$ = 0$^{\circ}$ and 90$^{\circ}$.  The amplitudes are normalized to the values at 0~T and at 
      the same $\theta$ to determine their relative change as the field is increased.
      (Right)  The relative timing resolution versus magnetic field.  A small horizontal shift was 
      added to the data at the same magnetic field so that the points are easier to see.
      \label{fig:H6614-70_35deg}}
  \end{center}
\end{figure}

In actual operating conditions, the orientation of the PMT ($\theta$) can vary between 0$^{\circ}$ and 
180$^{\circ}$ with respect to the magnetic field.  All previous studies were done within 
(0$^{\circ}$, 90$^{\circ}$)~\cite{Takasaki:1985,Enomoto:1993xz,Iijima:1996uc,Bonesini:2006zz,Bonesini:2007zz,Boeglin:2008,Bonesini:2012wu}.  
In our measurements, a systematic study was conducted by rotating the 2~in.\ PMT assembly to angles of 
215$^{\circ}$ and 325$^{\circ}$ (or $-$35$^{\circ}$), where 215$^{\circ}$ was achieved 
by rotating the dark box by 180$^{\circ}$. A set of tests were also conducted by rotating the azimuthal
angle ($\phi$) of the PMT by $\approx$ 90$^{\circ}$ counterclockwise while keeping $\theta$ at 35$^{\circ}$.  
The results from this study are presented in Fig.~\ref{fig:H6614-70_35deg} with the relative amplitudes 
(left-side panel) and timing resolutions (right-side panel) plotted versus magnetic field.  The mesh inside 
the FM PMTs is not necessarily rotationally symmetric around the PMT's axis because each was assembled 
by hand~\cite{private_Hamamatsu}.  However, the relative amplitudes and timing resolutions are fairly 
consistent to better than a few percent for all angle combinations measured.  This indicates the FM PMT's 
performance to be nearly independent of $\phi$ and symmetric around $\theta$ = 180$^{\circ}$.

As noted in previous studies~\cite{Bonesini:2006zz,Bonesini:2007zz,Bonesini:2012wu}, the FM PMTs 
perform well up to a critical angle $\theta_{c}$.  From the measurements presented here, $\theta_{c}$ appears 
to be $>$ 45$^{\circ}$ for the H6614-70 assembly, and this FM PMT can operate well in magnetic fields up to  
1.5$-$1.6~T for relative orientation angles between 30$^{\circ}$ and 45$^{\circ}$.   By 50$^{\circ}$, both 
the relative amplitude and timing resolution begin to degrade, even at lower magnetic fields.

Each of the measurements consisted of about $7 \times 10^{5}$ events, resulting in a negligible statistical
uncertainty compared to the systematics.  From the sources of uncertainty previously discussed, we estimate 
a measurement uncertainty of 2.5$-$6\% on the gain, where the lower value is for fields less than 1~T, and the 
larger for {\bf B} at and above 1~T.

\subsection{One inch PMT assembly (H6152-70)}
\label{sec:1in_results}
\begin{figure}[hbtp]
  \begin{center}
    \includegraphics[width=0.49\textwidth]{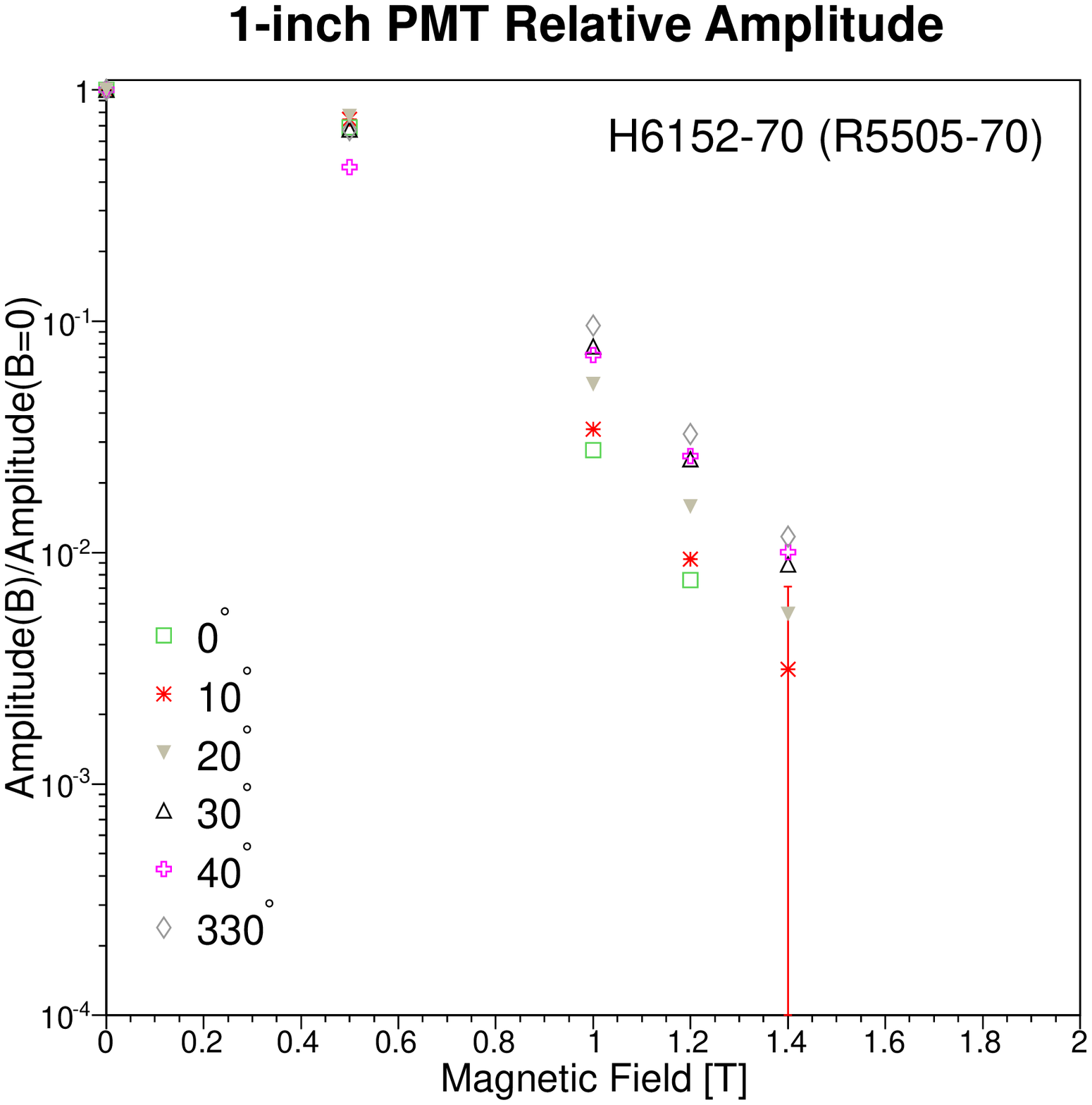}
    \includegraphics[width=0.49\textwidth]{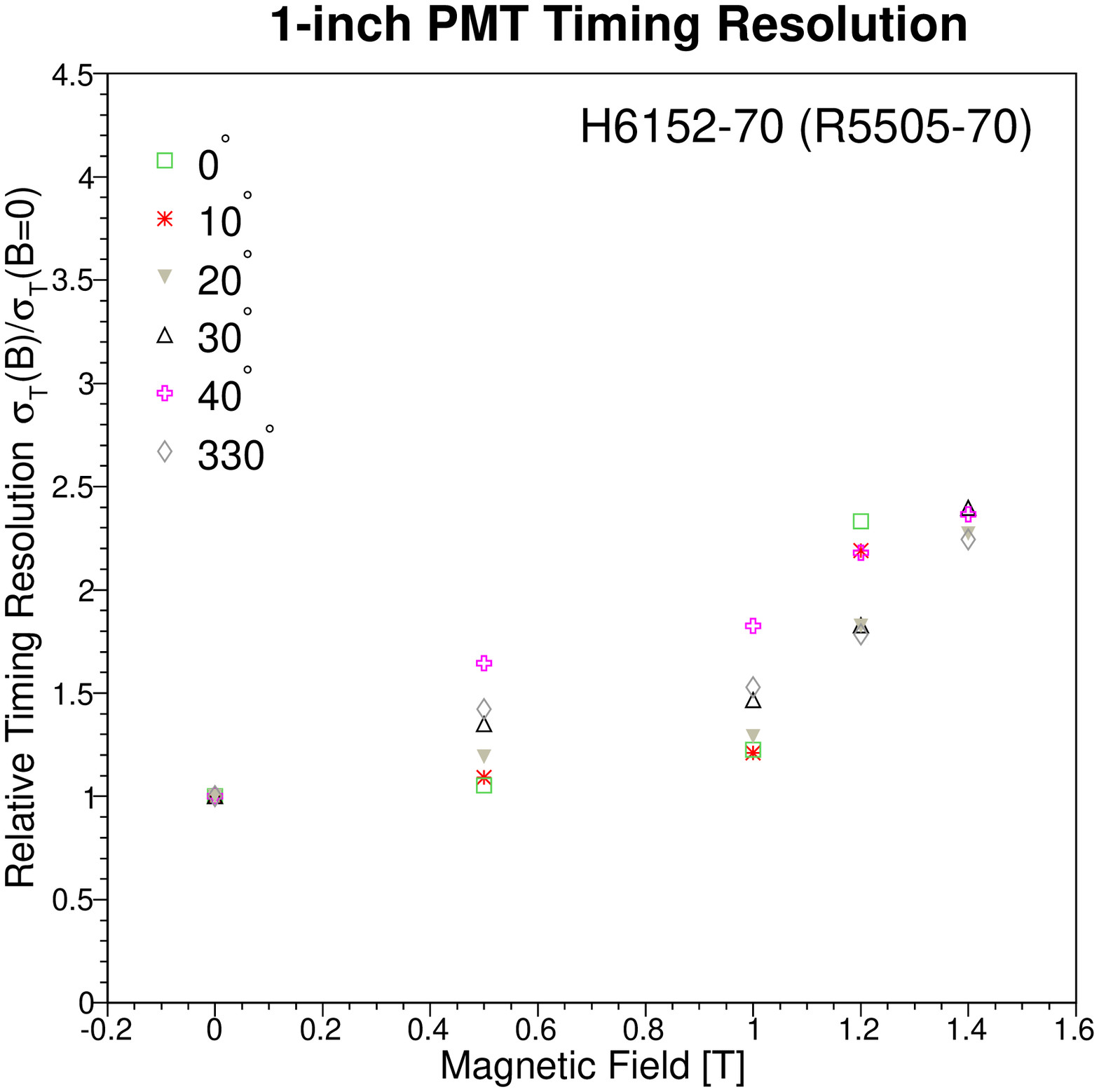}
    \caption{(Left) Relative amplitude versus magnetic field for the 1~in.\ Hamamatsu H6152-70 assembly
      at various $\theta$.   The amplitudes are normalized to the values at 0~T and at the same $\theta$
      to determine their relative change as the field is increased.   (Right) The relative timing resolution 
      versus magnetic field.
      \label{fig:H6152-70_all}}
  \end{center}
\end{figure}

For the H6152-70 (1~in.)~assembly, measurements were conducted for orientation angles from 0$^{\circ}$
up to 40$^{\circ}$ with magnetic fields from 0~T to 1.4~T.  At 0$^{\circ}$, the studies were conducted only 
up to 1.2~T due to a lack of signal in both the fADC and TDC at higher fields.  The relative amplitudes 
and timing resolutions versus magnetic field are presented in Fig~\ref{fig:H6152-70_all}.  Statistical error 
bars are plotted on the amplitudes, though in most cases they are too small to be visible.  The measurements 
for the 1~in.\ FM PMT demonstrated a worse performance in part due to the factor of 20 smaller gain at 2.0~kV 
compared to the 2~in.\ PMT.  The relative amplitude measurements have the same general behavior as for the 
larger PMT, but the decrease as the field increases is steeper.

For the H6152-70 assembly, $\sigma_{T}$ was also normalized to the results at $\theta$ = 0$^{\circ}$ 
and zero field and scaled to correct for the loss of $N_{PE}$ as $\theta$ is rotated away from 0$^{\circ}$.  
As the orientation angle increases, the timing resolution quickly worsens above 
20$^{\circ}$ and near magnetic fields of 1~T.  It was suspected that the smaller surface area of the 
photocathode (4 times) limited the timing resolution to 460~ps at 0$^{\circ}$ and zero magnetic 
field due to less light being collected by the FM PMT.  Measurements were taken by halving the 
distance from the diffuser to the PMT's surface from 20~cm to 10~cm, though no improvement was 
seen in either amplitude or timing resolution.   A dedicated systematic study is required to isolate 
the cause of the worse timing resolution for this FM PMT.   In order to test its 
symmetry with respect to the magnetic field direction, the FM PMT was also rotated to 330$^{\circ}$ 
(or $-$30$^{\circ}$).  The differences in the relative amplitudes were found to be $\approx$ 1\% and 
are within the reproducibility 
of these measurements and thus not systematically significant.  Error bars are not included on the timing 
resolution data to provide clarity in the figure.

From the measurements, it was found that the relative amplitude of the H6152-70 assembly improves with 
increasing $\theta$ up to 40$^{\circ}$ and up to 1.4~T.  However, the timing resolution worsens considerably 
at fields higher than 1~T and $\theta$ above 20$^{\circ}$.  The general trends seen in 
Ref.~\cite{Bonesini:2012wu} were reproduced, though the absolute timing resolution is 10 times worse in our 
measurements due to significantly fewer initial photoelectrons and the decay from the LED capacitance.  
Unfortunately, the pulse generator was already near the maximum amplitude, and there was no convenient way
to increase the amount of light incident on the photocathode.   The useful range of the H6152-70 appears to 
be smaller than the H6614-70, though a preamplifier with a higher gain could potentially extend its range 
to higher magnetic fields.

\section{Conclusions}
\label{sec:conclusions}
We measured the relative gain and the timing resolution of the H6152-70 and H6614-70 fine-mesh 
photomultiplier tube assemblies from Hamamatsu Photonics within a relative orientation angle ($\theta$) 
range between 0$^{\circ}$ and 50$^{\circ}$ and for a magnetic field range between 0 and 1.9~T.  The test 
results show that the 2~in.\ FM PMT (H6614-70 assembly) has a relative gain reduction of a factor of 100, 
while preserving good timing resolution for $\theta$ = 35$^{\circ}$ to 45$^{\circ}$ for fields up to 1.5~T.  
The 1-in.\ PMT (H6152-70 assembly) did not perform as well in either the relative amplitude and 
timing resolution results.  The timing resolution of the H6152-70 is especially sensitive to the 
orientation angle between the PMT axis and the field, and performs considerably worse above 20$^{\circ}$ 
and 1~T.  From these measurements, both FM PMTs show resiliency operating in moderate magnetic fields, 
though their effective ranges are different.  Our results also indicate that the FM PMT's performance is 
nearly independent of $\phi$ (azimuthal angle) and symmetric about $\theta$ = 180$^{\circ}$.  Even though 
our absolute timing resolution was limited to 300~ps, previous measurements in high magnetic fields up 
to 1.2~T have achieved resolutions better than 50~ps.  Based on all these measurements, the H6614-70 
assembly is found to be suitable for TOF measurements for the LASPD of SoLID (1.5~T field), where a 
timing resolution of 150~ps is required.  

\section{Acknowledgments}
The UVA collaboration wishes to acknowledge the DIRC Collaboration and Detector
Group at Jefferson Lab:  Brian Kross, Seung Joon Lee, Jack MacKisson, Pawel Nadel-Turonski 
and Carl Zorn for their support with the test facility and allowing us use of their system.  
We also would like to thank Sergey Boyarinov for support with implementation of the TDC 
into the readout, and the Hamamatsu representative Ardavan Ghassemi for useful 
discussions in preparing for these measurements.  This work was supported in part
by the U.S.\ Department of Energy under award no.\ DE-SC00014434 and by the Brookhaven 
National Laboratory under award eRD4.  This work was authored by Jefferson Science Associates, 
LLC under U.S.\ DOE contract no.\ DE-AC05-06OR23177.  The U.S.\ Government retains a nonexclusive, 
paid-up irrevocable, world-wide license to publish or reproduce this manuscript for U.S.\
Government purposes.




\bibliographystyle{elsarticle-num}

\section*{References}

\bibliography{fmpmt_NIM_paper}




\end{document}